\documentclass[cameraready]{Interspeech}
\usepackage[T1]{fontenc}
\title{ParA-LLM: A Unified Approach to \\ Paralinguistic and Acoustic Speech Understanding}
\author[affiliation={1,2}]{Nishit}{Anand}
\author[affiliation={1}]{Jiaqi}{Su}
\author[affiliation={1}]{Ke}{Chen}
\author[affiliation={1}]{Yunyun}{Wang}
\renewcommand{\authorsep}{,\\}
\author[affiliation={2}]{Dinesh}{Manocha}
\author[affiliation={2}]{Ramani}{Duraiswami}
\author[affiliation={1,3}]{Rithesh}{Kumar}
\author[affiliation={1}]{Zeyu}{Jin}

\address{
    $^1$ Adobe Research, USA $^2$ University of Maryland, College Park, USA $^3$ OpenAI, USA
}

\email{Corresponding email: nishit@umd.edu}

\keywords{paralinguistics, acoustics, speech understanding, large audio language models, large language models}

\usepackage{comment}

\usepackage{makecell}
\begin{document}

\maketitle

{%
\renewcommand{\thefootnote}{}
\footnotetext{\noindent Work done while Nishit Anand and Rithesh Kumar were at Adobe Research.}
}

% the abstract here must exactly match the abstract entered into the paper submission system
\begin{abstract}
    % 1000 characters. ASCII characters only. No citations.
    Recent advances in Audio LLMs have achieved human-level speech recognition, yet existing systems struggle to capture paralinguistic aspects such as speaker traits, expressive variations, and environmental acoustic conditions. To address this, we design a framework of 22 paralinguistic characteristics and create a dataset of over 1.2M Audio-QA pairs. We develop ParA-LLM, trained with a two-stage curriculum: first on single-attribute questions to build foundational knowledge, then on multi-attribute questions for joint reasoning over speaker and acoustic characteristics. We also release ParA-Bench, a benchmark of 6,000 multiple-choice questions across speaker-speech, acoustic, and mixed categories, where frontier models like GPT-4o-Audio achieve only 36\% accuracy. ParA-LLM surpasses state-of-the-art Audio LLMs like GPT-4o-Audio by 7.5\% on ParA-Bench, with additional gains of 1.13\% on MMAU-Pro Speech and 7.49\% on MMAR Speech.
\end{abstract}

\section{Introduction}
\label{sec:intro}
\begin{figure}[t]
  \centering
  \includegraphics[width=\linewidth]{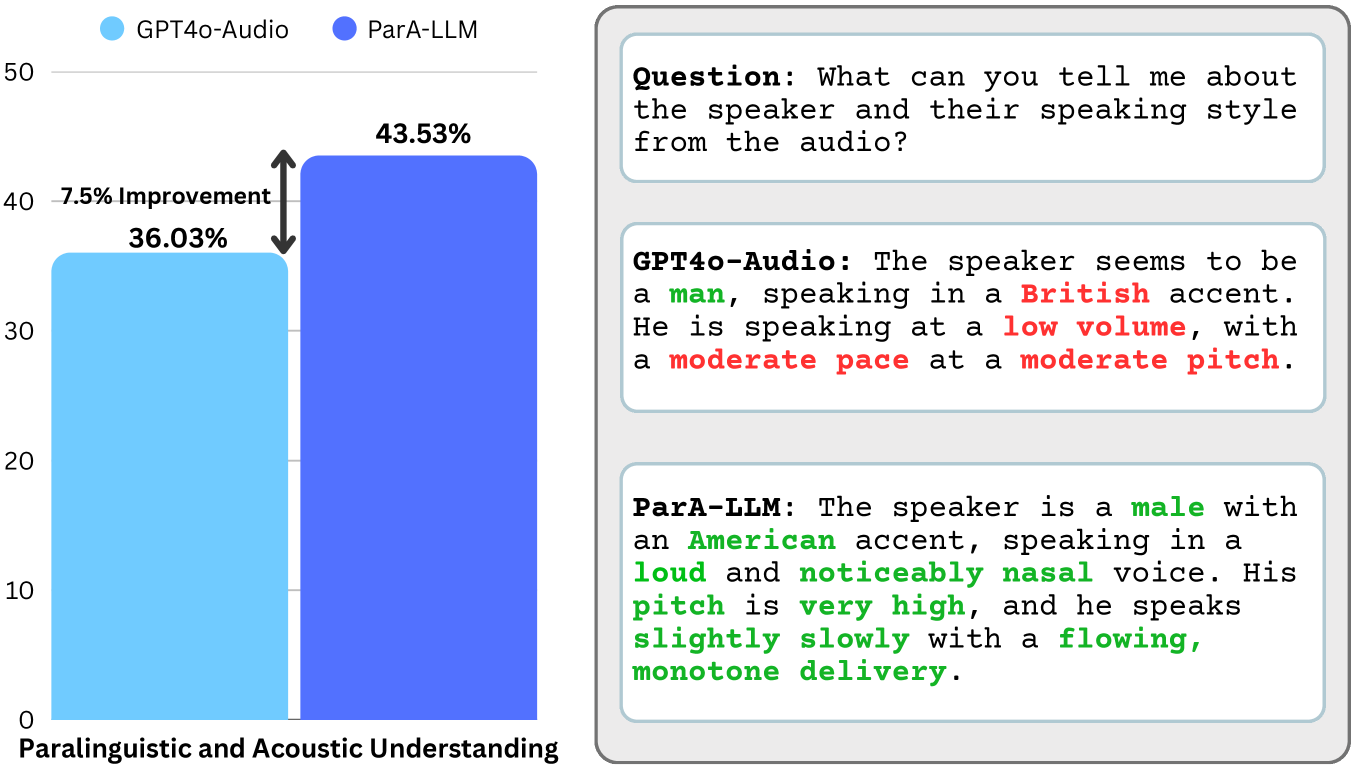}
  \caption{Left: Overall accuracy of paralinguistic and acoustic understanding across GPT-4o-Audio and ParA-LLM on ParA-Bench. Right: a QA example, where GPT-4o-Audio misidentifies speaker and speech characteristics.}
  \label{fig:hero_diagram}
  \vspace{-3mm}
\end{figure}

Recent advances in Audio LLMs~\cite{af3, voxtral} have led to significant progress in understanding the semantic and verbal content of speech. These models achieve strong performance on tasks such as automatic speech recognition (ASR), speaker diarization, and speech-to-speech translation. In particular, state-of-the-art systems demonstrate near-human performance in transcribing speech and reasoning over its verbal content, enabling them to answer complex queries about \textit{what was said}. For example, in ASR, humans achieve 97\% accuracy while state-of-the-art models reach 94\%, effectively closing the gap~\cite{asr-human-model}. However, as shown in Fig.~\ref{fig:hero_diagram}, despite these advances, existing models have limited understanding of paralinguistic aspects of speech, such as acoustic conditions, speaker traits, and expressive characteristics, which play an important role in natural human communication.

\noindent Prior literature~\cite{mmaupro} reports human accuracy of 82.3\% on common benchmarks like MMAU-Pro, while the best models perform well below 60\%. We made similar observations in a small controlled study using 100 paralinguistic questions from our benchmark: 78\% human accuracy versus 36\% by the state-of-the-art GPT-4o-Audio.

\noindent Paralinguistic understanding remains a largely open research problem.
We aim to bridge the gap by introducing data strategies and models that explicitly capture \textit{how something is being said, not just what is being said}.

\begin{figure*}[t]
  \centering
  \includegraphics[width=\textwidth]{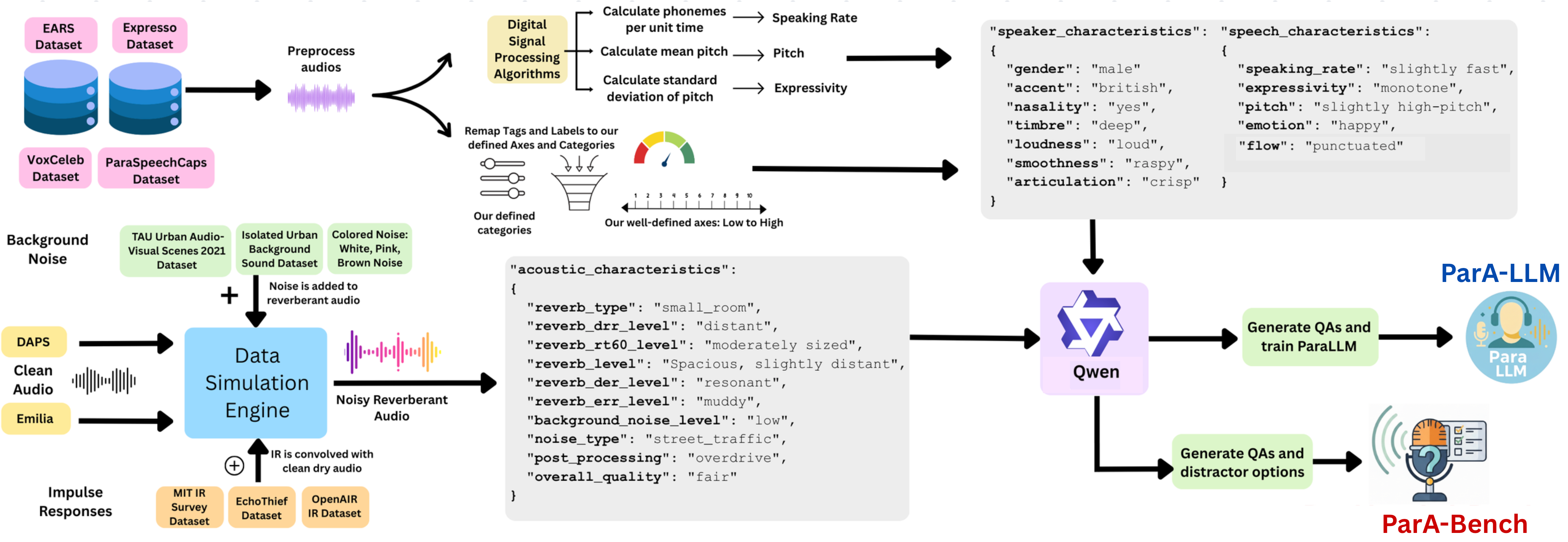}
    \caption{Our end-to-end pipeline for paralinguistic and acoustic understanding. \textbf{Data Simulation}: Clean speech is augmented with noise and reverberation for diverse acoustic conditions. \textbf{Characteristic Annotation}: Metadata is mapped to 22 well-defined characteristics across acoustic, speaker, and speech properties. \textbf{QA Generation}: Template-based and LLM-based ICL generation yield 1.2M audio--QA pairs across two curriculum stages. \textbf{Model Training}: ParA-LLM is trained on atomic then multi-attribute QA pairs. \textbf{Benchmark Creation}: ParA-Bench comprises 6K multiple-choice questions across speaker-speech, acoustic, and mixed categories.}
  \label{fig:main_diagram}
  \vspace{-3mm}
\end{figure*}

\noindent To this end, we introduce a structured taxonomy of 22 paralinguistic characteristics spanning acoustic, speaker-intrinsic, and speech-dynamic properties, and curate a large-scale dataset of over 1.2M Audio-QA pairs. We develop ParA-LLM, an audio LLM trained with a two-stage curriculum: first on 688K atomic single-attribute QA pairs to build foundational paralinguistic knowledge, then on 513K multi-attribute QA pairs to develop joint reasoning across acoustic, speaker, and speech characteristics. Unlike prior systems that address individual paralinguistic attributes in isolation~\cite{voxprofile}, ParA-LLM supports free-form question answering over multiple interacting attributes simultaneously. Our two-stage curriculum not only improves paralinguistic understanding but also benefits broader audio understanding beyond our benchmark, achieving 1.13\% improvement on MMAU-Pro Speech and 7.49\% on MMAR Speech over the base model. We further introduce ParA-Bench,\footnote{\href{https://nishitanand.github.io/paralinguistic-understanding-llm}{https://nishitanand.github.io/paralinguistic-understanding-llm}} a benchmark of 6,000 multiple-choice questions for comprehensive evaluation of paralinguistic and acoustic understanding.

\noindent Our \textbf{main contributions} are as follows:
\begin{enumerate}
    \item We introduce a \textbf{large-scale dataset} of over 1.2M Audio-QA pairs with structured annotations across 22 paralinguistic characteristics, and a two-stage curriculum training paradigm that progressively builds paralinguistic understanding from atomic to joint multi-attribute reasoning.
    \item We develop \textbf{ParA-LLM}, an audio LLM with unified paralinguistic and acoustic understanding, supporting free-form joint question answering across acoustic, speaker, and speech characteristics, surpassing state-of-the-art Audio LLMs by \textbf{7.5\%} on ParA-Bench.
    \item We release \textbf{ParA-Bench}, a 6,000 multiple-choice question benchmark to comprehensively evaluate models' ability to understand and reason over paralinguistic and acoustic characteristics, where frontier models achieve \textbf{36\% accuracy}.
\end{enumerate}

\section{Related Work}
\label{sec:related_work}
Recent years have seen rapid progress in Audio LLMs. Early systems such as Pengi~\cite{pengi} and LTU-AS~\cite{ltu-as} demonstrated the first steps in audio understanding, followed by GAMA~\cite{gama} and Audio Flamingo 3~\cite{af3}, which improved audio comprehension and free-form QA. More recently, reasoning-oriented models like R1-AQA~\cite{r1aqa}, Mellow~\cite{mellow}, and Audio-CoT~\cite{audio-cot} applied GRPO and chain-of-thought prompting to enhance complex reasoning, while general-purpose models such as Qwen2-Audio~\cite{qwen2audio} and Qwen2.5-Omni~\cite{qwen-omni} have shown strong results on content-based speech benchmarks. Despite this progress, these systems largely excel at verbal tasks, i.e.,\ \textit{what was said}, while lacking robust understanding of paralinguistic aspects such as acoustics (reverberation, noise, echo), speaker traits (timbre, articulation, roughness), and speech dynamics (speaking rate, pitch). Specialized models for emotion recognition, accent classification, or gender prediction address individual attributes in isolation, and cannot generalize across properties. Vox-Profile~\cite{voxprofile} trains separate classifiers covering a broad set of speaker and speech traits, but remains limited to single-attribute classification rather than joint reasoning over multiple interacting paralinguistic properties. In contrast, ParA-LLM supports free-form question answering over multiple attributes simultaneously, enabling compositional reasoning across acoustic, speaker, and speech-dynamic characteristics.

\noindent Curriculum learning~\cite{bengio2009curriculum} improves generalization by ordering training examples from simple to complex, and has been applied in speech and language domains to progressively build task understanding~\cite{lotfian2019curriculum, wijngaard2025curriculum}. Our model ParA-LLM is trained with a two-stage curriculum - first on atomic single-attribute questions to build foundational paralinguistic knowledge, then on complex multi-attribute questions to develop combined reasoning over speech and acoustic characteristics - enabling it to understand 22 paralinguistic and acoustic characteristics and support free-form question answering across multiple attributes at once.

\section{Methodology}
\label{sec:methodology}
We aim to build a model that understands acoustic, speaker, and speech properties. To this end, we define 22 paralinguistic characteristics covering acoustic and speaker-centric aspects of speech, including 10 acoustic characteristics, 7 speaker-intrinsic characteristics, and 5 utterance-level speech characteristics. Unlike prior work that often relies on subjective labels~\cite{dreamvoice}, our characteristics are designed to be distinct and objective wherever possible: acoustic properties use signal-based metrics (e.g., DRR, RT60, SNR), while speaker and speech attributes use natural-language descriptors with majority annotator agreement, following prior datasets such as ParaSpeechCaps~\cite{paraspeechcaps}. Each property is defined along a continuous axis (like speaking rate, noise level), binary attribute (like nasality), or multiclass category (like articulation, flow), ensuring consistency and interpretability. For all audio samples, we construct metadata records spanning acoustic, speaker, or speech characteristics, forming the foundation for data curation and QA generation. Fig.~\ref{fig:main_diagram} shows our full pipeline, including data simulation, annotation, QA generation, model training, and benchmark creation.

\noindent \textbf{Data Engine.} For acoustic characteristics, we utilize clean speech recordings from EARS~\cite{richter2024ears}, Emilia~\cite{emilia}, Expresso~\cite{nguyen2023expresso}, and VoxCeleb~\cite{nagrani2017voxceleb}. We develop an acoustic simulation engine which augments speech recordings with environmental effects by convolving them with room impulse responses (RIRs) and mixing in background noise samples at different signal-to-noise ratios (SNRs). As such, the data engine produces real-world-like acoustic scenes, with varying room sizes and geometries, wall surface materials, speaker distances, and ambient noises. Finally, the engine applies post-production effects such as clipping, dynamic range compression, and overdrive, further diversifying the acoustic conditions.

\noindent \textbf{Ten Acoustic Characteristics.} 
Measured room impulse responses (RIRs) are sourced from MIT IR Survey~\cite{mit_ir_survey} and EchoThief~\cite{echothief}, spanning diverse environments including classrooms, bedrooms, supermarkets, tunnels, and train stations, which are grouped into 15 defined \textbf{Reverb Types}. 

\noindent Meanwhile, we draw ambient noise samples from the TAU Urban Audio-Visual Scenes 2021~\cite{noise-tau} and Isolated Urban Sound Background~\cite{isolated-urban-sound}, supplemented with synthetic colored noises (white, pink, and brown), collectively mapped into 21 defined \textbf{Noise Types} and mixed at varying SNRs. Both reverb and noise types are defined based on real-world scenes, with acoustically similar environments consolidated into the same class.

\begin{table}[t]
  \caption{Accuracy of models on ParA-Bench across speaker-speech, acoustic, and mixed categories. \textbf{Bold} denotes the best and \underline{underline} denotes the second-best in each category.}
  \label{tab:main_results}
  \centering
  \small
  \setlength{\tabcolsep}{4pt}
  \begin{tabular}{lcccc}
    \toprule
    \textbf{Model} &
    \makecell{\textbf{Speaker-}\\\textbf{Speech}} &
    \textbf{Acoustic} &
    % \makecell{\textbf{Mixed}\\\textbf{(Acoustic \&}\\\textbf{Speaker-Speech)}} &
    \makecell{\textbf{Mixed}} &
    \textbf{Overall} \\
    \midrule
    Qwen2-Audio          & 34.80 & 23.45 & 25.90 & 28.05 \\
    Voxtral              & \underline{51.95} & 29.20 & \underline{35.25} & \underline{38.80} \\
    Audio Flamingo 3     & 34.35 & \underline{38.75} & 33.40 & 35.50 \\
    \midrule
    Mellow               & 3.60  & 3.60  & 3.80  & 3.67  \\
    R1-AQA               & 26.60 & 19.00 & 21.20 & 22.27 \\
    \midrule
    Qwen2.5 Omni 3B      & 20.85 & 9.45  & 10.35 & 13.55 \\
    Qwen2.5 Omni 7B      & 21.30 & 11.75 & 10.35 & 14.47 \\
    GPT-4o-Audio          & 32.00 & \textbf{41.85} & 34.25 & 36.03 \\
    \midrule
    % \hdashline
    \textbf{ParA-LLM (Ours)} & \textbf{55.85} & 34.80 & \textbf{39.95} & \textbf{43.53} \\
    \bottomrule
  \end{tabular}
  \vspace{-3mm}
\end{table}

\noindent The acoustic environment is further characterized using standard metrics and customized ones derived from the RIRs, each discretized into three levels by default. 
(1) \textbf{Reverb DRR Level} (Direct-to-Reverberant Ratio) captures the relative contribution of direct versus reverberant sound, reflecting microphone–speaker distance, and is discretized into three levels: \textit{close, moderately close, and distant}. 
(2) \textbf{Reverb RT60 Level} (Reverberation Time) quantifies the time for sound energy to decay by 60 dB, capturing perceived room spaciousness, and is discretized into \textit{tight, medium-sized, and spacious}. 
(3) We further derive the \textbf{Overall Reverb Level} using a 3×3 grid of DRR and RT60 bins, representing overall reverberation strength.
To analyze reverberation in finer granularity, we also include
(4) \textbf{Reverb ERR Level} (Early-to-Reverb Ratio, also known as C50) quantifies early energy beneficial for speech intelligibility relative to late-arriving energy beyond 50 ms; and lastly (5) \textbf{Reverb DER Level} (Direct-to-Early-Reflections Ratio) measures the energy ratio between the direct signal and early reflections, indicating whether speech sounds clear or resonant. 

\noindent In parallel, \textbf{Background Noise Level} categorizes the SNR into \textit{low, medium, or high} levels. \textbf{Post Processing} indicates whether effects such as clipping, dynamic range compression, overdrive, or phaser were applied during simulation. Finally, \textbf{Overall Quality} is quantified using STOI~\cite{taal2011stoi}, a standard speech intelligibility metric, discretized into three bins: \textit{poor} ($<$0.8), \textit{fair} (0.8–0.92), and \textit{high} ($>$0.92), determined empirically based on perceptual judgments by audio experts.

\noindent\textbf{Seven Speaker Characteristics.}
We consider speaker-intrinsic characteristics that remain consistent within a speaker but vary across individuals: \textbf{gender} (male/female), \textbf{accent} (multiclass), \textbf{nasality} (binary: nasal or non-nasal), \textbf{timbre} (multiclass: deep, shrill, etc.), \textbf{loudness} (multiclass: soft, hushed, loud, booming, etc.), \textbf{smoothness} (multiclass: silky, husky, raspy, etc.), and \textbf{articulation} (multiclass: crisp, enunciated, slurred, etc.). Audio is sourced from EARS~\cite{richter2024ears}, Emilia~\cite{emilia}, Expresso~\cite{nguyen2023expresso}, ParaSpeechCaps~\cite{paraspeechcaps} and VoxCeleb~\cite{nagrani2017voxceleb}. Although some datasets provide labels for attributes like gender or timbre, the annotations are often subjective and inconsistently defined across corpora. To ensure uniformity, we apply our definitions of axes and categorical mappings for each characteristic.

\noindent\textbf{Five Speech Characteristics.}
Unlike speaker traits, speech characteristics vary across utterances for the same speaker. We define five attributes: \textbf{Emotion}, spanning basic (happy, angry, sad, scared) and complex (enthusiastic, calm, anxious, confused, bored) categories; \textbf{Speaking Rate}, measured as phonemes per unit time and discretized into seven bins; \textbf{Pitch}, gender-adjusted and binned into seven levels; \textbf{Expressivity}, capturing intonational variation via pitch standard deviation, categorized into five bins; and \textbf{Flow}, a multiclass measure of fluency and rhythm. Labels from existing corpora are mapped to our defined axes and categories for consistency.

\noindent After this stage, we obtain over 700K unique audio samples with corresponding metadata annotations. We apply class balancing and selection criteria to ensure uniform distribution across characteristic categories. From this pool, 306K samples are selected for Stage 1 training and 217K for Stage 2 training, with a held-out test set of 6K samples strictly reserved for evaluation. The test set audio clips, RIRs, and noise samples are strictly disjoint from the training data, as manually verified.

\noindent \textbf{QA Generation for Training ParA-LLM.}
We train ParA-LLM using a two-stage curriculum. In \textbf{Stage 1}, we generate 688K atomic QA pairs across 306K unique audio samples using template-based generation, where each question targets exactly one paralinguistic or acoustic attribute (e.g., ``What is the gender of the speaker?''). This stage builds foundational, attribute-specific knowledge across all 22 characteristics. In \textbf{Stage 2}, we use Qwen2.5-7B~\cite{qwen7b} with in-context learning (ICL) to generate 513K multi-attribute QA pairs across 217K audio samples, where each question jointly queries multiple attributes (e.g., gender, accent, articulation, speaking rate, reverb, noise). This stage develops compositional reasoning over combined speaker, speech and acoustic characteristics.

\noindent \textbf{Creation of ParA-Bench.}
We introduce ParA-Bench, a 6K multiple-choice question benchmark spanning all 22 paralinguistic and acoustic characteristics, sampled from the held-out test set. Questions are evenly distributed across three categories: \textbf{speaker-speech} (2,000 questions on speaker-intrinsic and utterance-level speech characteristics), \textbf{acoustic} (2,000 questions on acoustic environment characteristics), and \textbf{mixed} (2,000 questions jointly querying both acoustic and speaker-speech characteristics). All questions are multi-attribute. To mitigate self-referential bias, ParA-Bench QA pairs are generated using Mistral-Small-3.2~\cite{mistral2025small32}, a different model family from the one used for generating training QAs (Qwen2.5-7B), ensuring no generative bias is shared between train and test sets. Distractor options for each question are generated by re-prompting the model to produce three plausible but incorrect alternatives. Human verification is further conducted on a subset of 300 questions to ensure benchmark reliability.

\section{Experimental Setup}
\noindent\textbf{Baselines.} 
We evaluate a wide range of models on ParA-Bench to assess their ability to capture acoustic, speaker, and speech characteristics, including state-of-the-art Audio LLMs such as GPT-4o-Audio~\cite{openai2024gpt4ocard}, Qwen2-Audio~\cite{qwen2audio}, Voxtral-24B~\cite{voxtral}, Audio Flamingo 3~\cite{af3}. We also test large audio reasoning models, including Mellow~\cite{mellow} and R1-AQA~\cite{r1aqa}, which incorporate explicit reasoning mechanisms. To examine performance in multi-modal settings, we further evaluate Omni models, including Qwen2.5-Omni 3B and 7B~\cite{qwen-omni}.

\noindent\textbf{Human Verification.}
To evaluate automatic judging, we sampled 100 Qwen2-Audio responses on ParA-Bench and had human annotators label them as correct or incorrect. The same set was then judged by three methods: an embedding-based approach (NV-Embed-v2~\cite{nvembed}, choosing the option with highest cosine similarity), an LLM-as-judge using Gemma3-27B~\cite{gemma3} (matching the response to the closest option), and a regex-based matcher (extracting A/B/C/D). Agreement with human labels showed LLM-as-judge achieved the best alignment at 91\%, compared to 68\% for embeddings and 43\% for regex.

\noindent\textbf{ParA-LLM Model training.} 
We initialize ParA-LLM from Qwen2-Audio-7B-Instruct~\cite{qwen2audio} and train it using a two-stage curriculum with LoRA (rank 128, alpha 256, dropout 0.1) applied to the audio encoder, multimodal projector, and LLM. In \textbf{Stage 1}, the model is trained on 688K atomic QA pairs for 1 epoch with a learning rate of 5e-5. In \textbf{Stage 2}, the Stage 1 LoRA adapter is loaded and training continues on 513K multi-attribute QA pairs for 1 epoch with a learning rate of 4e-5. Both stages use a cosine learning rate scheduler and AdamW optimizer. Training is conducted on 8 A100 GPUs with a per-device batch size of 16, giving a total batch size of 128.

\begin{table}[t]
  \caption{Performance of ParA-LLM trained with curriculum learning compared to the Qwen2-Audio-7B-Instruct baseline, demonstrating steady improvement across the benchmarks.}
  \label{tab:curriculum}
  \centering
  \scriptsize
  \setlength{\tabcolsep}{3pt}
  \begin{tabular}{lcccc}
    \toprule
    \textbf{Model} &
    \makecell{\textbf{MMAU-Pro}\\\textbf{(Speech)}} &
    \makecell{\textbf{MMAR}\\\textbf{(Speech)}} &
    \makecell{\textbf{MMAR}\\\textbf{(Sound-Speech)}} &
    \makecell{\textbf{MMAR}\\\textbf{(Overall)}} \\
    \midrule
    Qwen2-Audio-7B-Instruct         & 40.96 & 35.37 & 40.83 & 36.00 \\
    \midrule
    ParA-LLM Stage 1  & 41.98 & 37.76 & 45.87 & 39.40 \\
    ParA-LLM Stage 2  & \textbf{42.09} & \textbf{42.86} & \textbf{46.33} & \textbf{39.70} \\
    \bottomrule
  \end{tabular}
  \vspace{-3mm}
\end{table}

\section{Results and Analysis}

\noindent\textbf{Performance on ParA-Bench.} Table~\ref{tab:main_results} shows ParA-LLM achieves the highest speaker-speech accuracy at 55.85\% among all the models as well as for "mixed" and "overall", while GPT-4o-Audio leads on acoustic accuracy at 41.85\%. Overall ParA-LLM achieves 43.53\% accuracy, winning a margin of 4.73\% from the second-best model Voxtral at 38.80\%.

\noindent\textbf{Unimodal vs. Multimodal Models.}
To examine whether multi-modality benefits paralinguistic understanding, we evaluate omni models including Qwen2.5-Omni 3B and 7B. Both models perform poorly across all categories, with Qwen2.5-Omni 3B scoring as low as 9.45\% on acoustic and 13.55\% overall. These results suggest that capabilities with other modalities do not directly translate to improved generalization of paralinguistic and acoustic understanding.

\noindent\textbf{Audio LLMs vs. Audio Reasoning Models.}
Reasoning-based models such as R1-AQA and Mellow, which incorporate GRPO and chain-of-thought prompting, do not display consistent improvements over conventional audio LLMs. Mellow achieves only 3.67\% overall and R1-AQA reaches 22.27\%, indicating that current reasoning mechanisms do not directly enhance paralinguistic and acoustic understanding capabilities.

\noindent\textbf{Effect of Two-Stage Curriculum Learning.}
Table~\ref{tab:curriculum} shows the effect of our two-stage curriculum on standard audio benchmarks. Starting from the Qwen2-Audio-7B-Instruct baseline, Stage 1 training on atomic single-attribute QA pairs already shows consistent gains across all benchmarks, improving MMAU-Pro Speech from 40.96\% to 41.98\% and MMAR Speech from 35.37\% to 37.76\%. Stage 2 training on multi-attribute QA pairs further improves performance, reaching 42.09\% on MMAU-Pro Speech and 42.86\% on MMAR Speech: a 7.49\% absolute gain over the baseline on MMAR Speech and 3.70\% on MMAR Overall. These results demonstrate that our curriculum not only improves paralinguistic understanding on ParA-Bench, but also generalizes to broader audio benchmarks, with each stage contributing incrementally to the overall gains.

\section{Downstream Applications of ParA-LLM}
ParA-LLM’s deep knowledge of paralinguistics and acoustics enables a broad range of downstream applications. 

\noindent First, it can facilitate scalable, rich data annotation by automatically generating detailed perceptual metadata together with descriptive captions for speech corpora.

\noindent Second, it can infer fine-grained speech attributes such as articulation, timbre, emotion, and smoothness, providing structured controls for text-to-speech generation. Moreover, ParA-LLM supports emerging agentic speech editing workflows by answering targeted queries, assisting downstream agents to plan and execute editing operations.

\noindent As a concrete example, we demonstrate a novel application, text-to-impulse response generation (Text2IR), enabled by the acoustic understanding capability of ParA-LLM. We prompt ParA-LLM to produce detailed natural language descriptions of target acoustic spaces corresponding to room impulse responses, yielding $\sim$150K IR–caption pairs. Gencho~\cite{lin2025gencho} leverages this generated IR-caption dataset to train a diffusion-based text-conditioned IR generator. Its quantitative analysis shows the generated room impulse responses closely follow the semantics of text prompts, with examples online.

\section{Conclusion}

Paralinguistic understanding - capturing \textit{how} speech is delivered rather than simply \textit{what} is said - remains a fundamental gap in current audio AI systems. In this work, we presented a comprehensive framework to address this gap. We introduced a structured taxonomy of 22 paralinguistic and acoustic characteristics, and built a large-scale dataset of over 1.2M audio--QA pairs through a combination of acoustic simulation, systematic annotation, and both template-based and LLM-driven generation. Building on this foundation, we developed ParA-LLM, trained via a two-stage curriculum that advances from foundational single-attribute reasoning to complex multi-attribute joint inference over speaker, speech, and acoustic properties.

\noindent Our evaluations reveal that paralinguistic understanding remains a substantially unsolved problem: even frontier models such as GPT-4o-Audio achieve only 36\% accuracy on ParA-Bench, compared to human accuracy of 78\%, highlighting a wide gap. ParA-LLM closes a significant portion of this gap, surpassing GPT-4o-Audio by 7.5\% overall. Importantly, the gains are not confined to our benchmark: the curriculum training yields consistent improvements on MMAU-Pro Speech and MMAR Speech, demonstrating broad generalization. We further showed that neither multi-modal omni models nor explicit chain-of-thought reasoning mechanisms provide meaningful improvements on paralinguistic tasks, underscoring the importance of domain-specific approaches.

\noindent Finally, we demonstrated the practical utility of ParA-LLM through downstream applications including automatic speech annotation, fine-grained TTS control, and acoustic space captioning for text-conditioned impulse response generation. We release ParA-Bench, ParA-LLM, and the full dataset to support future research in this underexplored but critical dimension of speech understanding.

\section{Generative AI Use Disclosure}
In the preparation of this manuscript, AI-based tools were used in a restricted, well-defined capacity, limited to proofreading tasks such as fixing grammar, and polishing the writing. All technical content, including the research methodology, data analysis, and conclusion, was developed entirely without AI assistance.

\bibliographystyle{IEEEtran}
\bibliography{mybib}

\end{document}